\documentclass[prl,aps,twocolumn,10pt,nofootinbib,floatfix]{revtex4}
\usepackage{amsfonts}
\usepackage{amsmath}
\usepackage{amssymb}
\usepackage{epsfig}
\usepackage{slashed}
\usepackage{graphicx}
\usepackage{color}

\newcommand{\be}{\begin{equation}}
\newcommand{\ee}{\end{equation}}
\newcommand{\ba}{\begin{eqnarray}}
\newcommand{\ea}{\end{eqnarray}}

\newcommand{\pvec}{{\bf p}}

\newcommand{\dtilde}[1]{\frac{d^3 #1}{(2\pi)^3}}

\begin{document}
\title{Equation of state of a quark---Polyakov loop---meson mixture \\ in the PNJL model at finite temperature}
\author{Juan M. Torres-Rincon}
\author{Joerg Aichelin}
\affiliation{Subatech, UMR 6457, IN2P3/CNRS, Universit\'e de Nantes, \'Ecole de Mines de Nantes, 4 rue Alfred Kastler 44307,
Nantes, France}
\begin{abstract} 

  Recent consensus on the $N_f=2+1$ equation of state at vanishing chemical potential from different lattice-QCD groups
has spoiled the previous agreement with the outcome from the mean-field Polyakov-Nambu-Jona-Lasinio model. In this letter we
review the thermodynamics of the PNJL model introducing two important aspects needed to describe the pressure computed
in the lattice QCD. First, we consider the thermodynamics of the model beyond the mean-field approach to include pseudoscalar and scalar mesonic-like fluctuations into the grand-canonical potential.
This accounts for the hadronic pressure of the system below the critical temperature. On the other hand we also implement the back reaction of quarks into the
Polyakov-loop effective potential bringing a reduction of the pressure above $T_c$ from the Stefan-Boltzmann limit. We get a good
agreement with lattice-QCD data at low and moderate temperatures, opening the door to a straightforward extension to finite chemical potential.

\end{abstract}
\maketitle

\section{Introduction}

    The Nambu--Jona-Lasinio (NJL) model~\cite{Nambu:1961tp,Vogl:1991qt,Klevansky:1992qe,Hatsuda:1994pi,Buballa:2003qv} 
has been extensively used in the context of strong interactions due to its ability to account for several key
phenomena of the Quantum Chromodynamics (QCD), like the spontaneous symmetry breaking (together with the generation of
Goldstone bosons) and its restoration at high temperatures and densities. This model works as an effective realization of 
QCD at low energies, and allows us performing studies in a much simpler way in the regime where QCD is too difficult
to solve, or computationally expensive like in lattice-QCD calculations. 

    The absence of dynamical gluons --which are integrated out of the model-- has been partially compensated by the coupling of the
Polyakov loop effective potential to the quark sector~\cite{Ratti:2005jh}. Static quantities like the equation of state (EoS) showed --already at mean-field level-- a nice agreement with lattice-QCD data from 
the Bielefeld group (later Hot QCD collaboration)~\cite{Allton:2003vx,Cheng:2009zi}. This accordance was not only achieved at zero chemical potential ($\mu_B$),
but also at finite baryonic density using the Taylor expansion technique in the lattice data~\cite{Ratti:2005jh,Roessner:2006xn,Friesen:2011wt}. The quantitative agreement was not only surprising due to the relatively
small number of parameters of the model (quark bare masses, fermion coupling, ultraviolet (UV) cutoff...), but also due to the fairly good description at high temperatures
(where dynamical gluons should start to contribute) and below the transition temperature $T_c$ (where hadronic contribution 
is expected to dominate)~\cite{Ratti:2005jh,Costa:2010zw,Friesen:2011wt}. 
However, it is important to remind that the EoS computed from lattice-QCD was not unique, existing alternative results from the Wuppertal--Budapest
group~\cite{Aoki:2006br}, whose EoS at $\mu_B=0$ was not compatible with the PNJL model~\cite{Marty:2013ita}. 

     Recently, a consensus on the EoS between the two collaborations has been achieved, solidly pointing to a crossover transition around $T_c \simeq 155$ MeV, and a
compatible EoS up to temperatures of 400 MeV~\cite{Borsanyi:2013bia,Bazavov:2014pvz}. The final EoS --the one from the Wuppertal--Budapest group-- now disfavors the results from
the PNJL model. The pressure of the PNJL model underestimates the lattice-QCD results below $T_c$ (demonstrating the lack of hadronic pressure), and presents a rapid increase towards
the Stefan-Boltzmann (SB) limit of massless quarks and gluons, overshooting the lattice-QCD pressure soon after $T_c$~\cite{Marty:2013ita}.

The goal of this letter is to readdress the thermodynamics of the PNJL model and show that it can still provide a good
description of lattice-QCD EoS, once two natural ingredients are taken into account. At low temperatures one expects
that the main pressure should be carried by hadrons, primarily by the lightest mesons. To account for these states
we compute mesonic-like fluctuations of the grand-canonical potential on top of the mean-field result. This improvement has been
explored before in the literature, following the pioneering works of Refs.~\cite{Quack:1993ie,Hufner:1994ma}. At moderate temperatures $T\sim(1-2)T_c$, and due to the fast
restoration of chiral symmetry, the PNJL model describes a gas of nearly massless quarks and gluons. We will improve the description in this window by
considering the backreaction of quarks onto the Polyakov-loop effective potential. This effect --so far not applied to the PNJL model-- carries a systematic reduction
of the ``gluonic'' pressure bringing down the total pressure from the SB limit. In this regime, we will follow the outcome of a recent paper~\cite{Haas:2013qwp} in which
this idea has been exploited in detail using functional renormalization group methods.

\section{PNJL model with glue effective potential}

  In this study we follow the same conventions and parameters of the $N_f=2+1$ PNJL model of our past work~\cite{Torres-Rincon:2015rma}. 
We consider the local version of the NJL model with scalar and pseudoscalar interactions among quarks. The NJL Lagrangian is supplemented by
the 't Hooft (six-quark) interaction responsible to mimic the presence of the axial anomaly in the mass gap between the $\eta$ and $\eta'$ mesons. 
Finally, we also include the Polyakov-loop effective potential accounting for static gluonic properties. The grand-canonical potential of the model is expressed as
\be \label{eq:Omega} \Omega_{ \rm{PNJL}} (T,\mu_i;\Phi,\bar{\Phi}) = \Omega_{q} (T,\mu_i)+ {\cal U} (T;\Phi,\bar{\Phi}) \ , \ee
where the first term corresponds to the quark sector and the second one to the Polyakov-loop effective potential. 
We denote the temperature by $T$, quark chemical potentials by $\mu_i=(\mu_u,\mu_d,\mu_s)$, the
expectation value of the Polyakov loop (EVPL) by $\Phi$, and its conjugate by $\bar{\Phi}$.

\begin{figure}[htp]
\begin{center}
\includegraphics[width=0.45\textwidth]{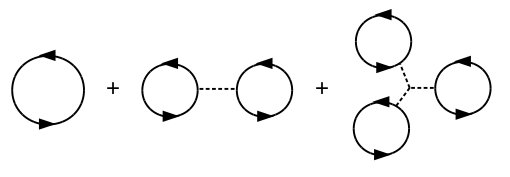}
\caption{\label{fig:Omega1} Noninteracting, Hartree and 't Hooft diagrams contributing at ${O} (N_c)$ to
the grand-canonical potential.}
\end{center} 
\end{figure}

  The grand-canonical potential for quarks will be expanded in inverse powers of the number of colors $N_c$~\cite{Hufner:1994ma}, 
\be \Omega_{q} (T,\mu_i)= \Omega_{q}^{(-1)} (T,\mu_i) + \Omega_q^{(0)} (T,\mu_i) + \cdots \ , \ee
where the index $(a)$ refers to the contribution at order of ${\cal O}( 1/N_c^a)$. The leading order term
(${\cal O}(N_c)$) contains the mean-field approximation, and correspond to the diagrams shown in Fig.~\ref{fig:Omega1} 
(where the quark interaction has been explicitly drawn with a dashed line). The explicit expression for $\Omega_{q}^{(-1)} (T,\mu_i)$
can be found in~\cite{Hansen:2006ee,Torres-Rincon:2015rma}.

    The quark propagators in the diagrams of Fig.~\ref{fig:Omega1} are taken in the Hartree+'t Hooft approximation, i.e. 
they are selfconsistently obtained from the minimization of $\Omega_q^{(-1)}$. This approximation is also consistent with 
a fixed order in $N_c$ counting. 

  In the absence of quarks --i.e. for a Yang-Mills theory-- the parameters of the Polyakov-loop 
effective potential ${\cal U}={\cal U}_{\rm{YM}}$ are chosen to fit the EoS of pure gauge lattice-QCD calculation, and to reproduce the EVPL 
as a function of temperature~\cite{Ratti:2005jh}. This procedure fixes the position of the minimum of ${\cal U}_{\rm{YM}}$ as a function of temperature,
but not the exact form of the effective potential itself. This freedom led to several parametrizations for ${\cal U}_{\rm{YM}}$ (``polynomial'', ``logarithmic''...),
all of them with a first-order deconfinement transition around $T_0=270$ MeV. As argued in Ref.~\cite{Schaefer:2007pw}, the effects of dynamical quarks bring down the
value of $T_0$ to a lower temperature. In Ref.~\cite{Torres-Rincon:2015rma} we took $T_0=190$ MeV for $N_f=2+1$ flavors,
consistent with the findings in~\cite{Schaefer:2007pw}. 

  Once the effective potential ${\cal U}_{\rm{YM}}$ is fixed and coupled to the NJL grand-canonical potential, the equilibrium configurations
of the order parameters are obtained by minimizing Eq.~(\ref{eq:Omega}). In this way, gluons modify the gap equations for quark masses (the Fermi occupation number is corrected by the EVPL).
In addition, the EVPL ($\Phi,\bar{\Phi})$ is also affected by quarks, transforming the deconfinement transition into a crossover around $T_0$. 
Notice that the Polyakov-loop effective potential itself is not modified by quarks, keeping its functional form, but with an EVPL which does not coincide 
with the minimum of ${\cal U}_{\rm{YM}}$ anymore.

  To compute the gluonic part of the pressure, $P=-{\cal U}_{\rm{YM}} (T;\Phi,\bar{\Phi})$, the effective potential is evaluated at the EVPL, probing a region of ${\cal U}_{\rm{YM}}$ not affected
by quark dynamics. Therefore, it is reasonable to ask for the effects of quarks into the effective potential itself, and not only 
at the minimum. Thus one considers not the pure gauge potential anymore, but the ``glue'' potential
 ${\cal U}_{\rm{glue}}$, which contains back reaction effects from dynamical quarks. A systematic study of this potential and its differences with respect to
 ${\cal U}_{\rm{YM}}$ has been made in the context of the functional renormalization group (FRG) in
Ref.~\cite{Haas:2013qwp}. We will follow the results of Ref.~\cite{Haas:2013qwp} and couple to the NJL Lagrangian
the ``glue'' potential ${\cal U}_{\rm{glue}}$. The FRG study shows that in a good approximation this potential is
related to the original YM potential by the transformation,
\be \label{eq:paw} \frac{{\cal U}_{\rm{glue}}}{T^4} (T;\Phi, \bar{\Phi}) = \frac{{\cal U}_{\rm{YM}}}{T^{*4}} 
(T^*;\Phi, \bar{\Phi}) \ , \ee
where $T^*$ depends on the temperature $T$ by
\be \frac{T^* - T^{*,cr}}{T^{*,cr}} = 0.57 \ \frac{T-T^{cr}}{T^{cr}} \ , \ee
with $T^{*,cr}=270$ MeV playing the role as the transition temperature in the YM case, and $180$ MeV $\lesssim T^{cr} \lesssim 270$ MeV, 
the one in the glue effective potential. The numerical coefficient 0.57 is the outcome from the comparison of both potentials within the FRG study. In the present work, 
we fix ${\cal U}={\cal U}_{\rm{glue}}$ setting $T_{cr}=190$ MeV. The parameters of the YM effective potential are taken as in~\cite{Torres-Rincon:2015rma}. We remove the UV cutoff 
from all the convergent integrals (this renormalization criterion is usually done with the unique aim of reaching the Stefan-Boltzmann limit in the quark sector at large temperatures).

  The grand-canonical potential at mean field, together with the glue potential can be used to solve the quark gap equations
and compare the values of the quark condensate with the lattice-QCD results. As the chiral transition temperature (inflection point of the quark condensate) appears in the PNJL shifted to higher temperatures, we plot 
the quark condensate as a function of $T/T_{c}$. The renormalized quark condensate $\Delta_{l,s}$ is defined as~\cite{Borsanyi:2010bp}
 \be \Delta_{l,s} (T) \equiv \frac{\langle \bar{q}q \rangle (T) - \frac{m_{q0}}{m_{s0}} \langle \bar{s}s\rangle (T) }
{\langle \bar{q}q \rangle (0) - \frac{m_{q0}}{m_{s0}} \langle \bar{s}s\rangle (0) } \ , \ee
 where $T_{c}=157 \pm 6 $ MeV in the lattice-QCD results of Ref.~\cite{Borsanyi:2010bp}. $q$ denotes any of the light quarks (we take them as degenerate).

\begin{figure}[htp]
\begin{center}
\includegraphics[width=0.42\textwidth]{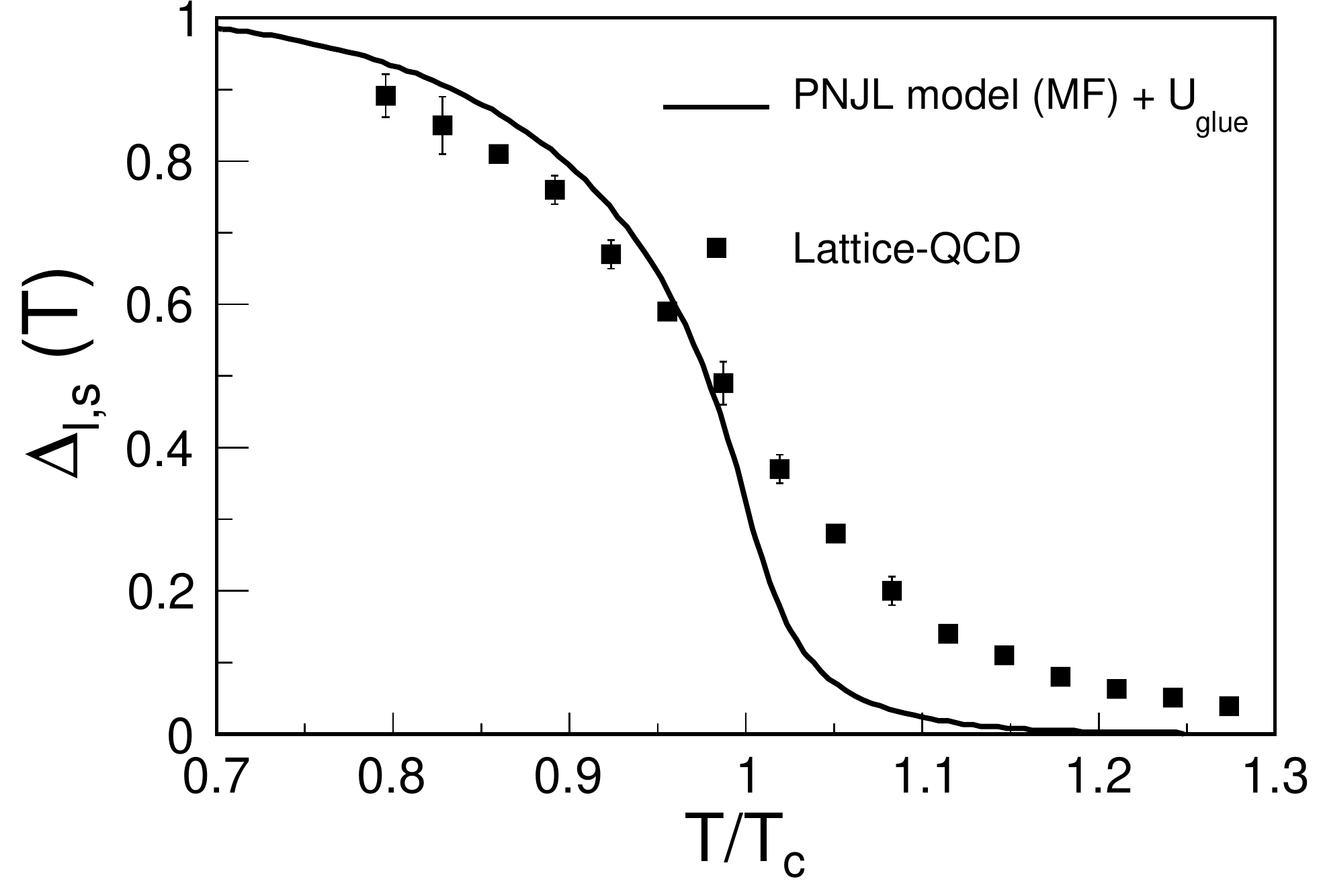}
\end{center} 
\caption{\label{fig:condlattice} Renormalized quark condensate as a function of the temperature versus $T/T_{c}$ in the PNJL model
computed at mean-field level. The dots correspond to the lattice-QCD calculation of Ref.~\cite{Borsanyi:2010bp}. }
\end{figure}

  In Fig.~\ref{fig:condlattice} we compare the result from the PNJL model with the lattice-QCD calculation of Ref.~\cite{Borsanyi:2010bp}.
In the PNJL model we compute $T_{c}$ as the inflection point of the quark condensate, which for
the light sector it reads $T_{c}=209$ MeV (for the strange one lies very close at $T_{c}=208$ MeV).
A similar trend for the renormalized condensate is captured in the Polyakov-loop quark-meson model
of Ref.~\cite{Haas:2013qwp}.

\section{\label{sec:fluc} Mesonic fluctuations}

\begin{figure}[htp]
\begin{center}
\includegraphics[width=0.45\textwidth]{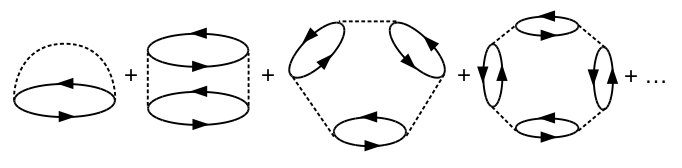}
\caption{\label{fig:Omega0} Fock and ring diagrams contributing to the grand-canonical potential
at ${\cal O} (N_c^0)$. Mesonic contributions to the pressure are encoded in the ring summation.}
\end{center} 
\end{figure}

   At low-temperatures the Polyakov loop contribution is able to suppress unphysical pressure from
quarks~\cite{Ratti:2005jh}. However, the mean-field computation at $O(N_c)$ cannot account for the hadronic pressure at $T<T_c$. 
The mesonic contribution to the pressure is encoded in the fluctuations over the mean field contained in $\Omega_q^{(0)}$~\cite{Quack:1993ie,Hufner:1994ma, Zhuang:1994dw,Blaschke:2007np}. 
At this order one should introduce the ``Fock diagram'' and the infinite set of ring diagrams depicted in Fig.~\ref{fig:Omega0}.

   The quark propagators computed from the diagrams of Fig.~\ref{fig:Omega1} are also used now for the diagrams of
Fig.~\ref{fig:Omega0}. Though one looses self-consistency, this approximation is still consistent with $N_c$ counting~\cite{Hufner:1994ma,Zhuang:1994dw,Oertel:2000cw}. 
From this perturbative addition, mesonic correlations emerge in the thermodynamical potential. This can also be seen
in terms of scattering amplitudes, where the equivalent approach is the solution of a Bethe-Salpeter equation 
for the quark-antiquark scattering in the ``random-phase approximation''. The resulting scattering amplitudes
contain poles in the complex energy plane associated to mesonic states. Masses and decay widths of these states
(associated to the real and imaginary parts of the pole position) were calculated for instance in Ref.~\cite{Torres-Rincon:2015rma}.

  The computation of $\Omega_q^{(0)}$ can be performed using the ``coupling-constant integration
technique''~\cite{Klevansky:1992qe,Hufner:1994ma,Zhuang:1994dw}. The total contribution to the thermodynamic potential
comes from every possible spin-flavor meson channel is
\be \Omega_q^{(0)} (T,\mu_i) = \sum_{M \in \pi, K, {\bar K}, \eta, \eta', f_0} \Omega^{(0)}_M (T,\mu_M (\mu_i))\ , \ee
where we have considered all possible scalar and pseudoscalar states with a vacuum mass below 1 GeV.
Notice that the meson chemical potential is a function of the quark chemical potential, $\mu_M=\mu_q - \mu_q'$,
with $q$ and $q'$ denoting the valence quark and antiquark of each meson.

  In the imaginary-time formalism, each individual contribution reads~\cite{Zhuang:1994dw,Hufner:1994ma,Blaschke:2013zaa,Wergieluk:2012gd}
\be \label{eq:omega0} \Omega^{(0)}_{M} (T,\mu_M)= \frac{g_M}{2} T \sum_n \int \dtilde{p} \log [ 1-2K_M\Pi_M (i \omega_n, \pvec) ] \ , \ee
where $g_M$ is the spin-isospin degeneracy factor for each meson, $K_M=K_M(T)$ is the quark coupling
(cf. Ref.~\cite{Torres-Rincon:2015rma} for details) and $\Pi_M(i\omega_n,{\bf p};T,\mu_i)$ is the $q-\bar{q}$ polarization function with the
appropriate quantum numbers of the state $M$. Explicit formulas are shown in Ref.~\cite{Rehberg:1995kh,Hufner:1994ma,Torres-Rincon:2015rma}.

We remind that one can connect the expression in Eq.~(\ref{eq:omega0}) with the so-called Beth-Uhlenbeck approach,
where the thermodynamic potential is expressed in terms of the $\bar{q}-q$ scattering phase-shift in the appropriate channel~\cite{Zhuang:1994dw,Hufner:1994ma,Blaschke:2013zaa,Wergieluk:2012gd}.
For this, one makes use of the Jost representation of the {\cal S}-matrix element, which is related to
the scattering phase-shift as $s_M (\omega, \pvec; T,\mu_M)= e^{2i\delta_M(\omega, \pvec; T,\mu_M)}$~\cite{taylor1972scattering}.
The latter is introduced as
\be \delta_M (\omega, \pvec; T,\mu_M)= - \textrm{ Arg } [1 - 2 K_M \Pi_M (\omega +i \epsilon, \pvec) ] \ , \ee
where we have performed the analytic continuation of the integrand in Eq.~(\ref{eq:omega0}) to real energies ($i\omega_n \rightarrow \omega+i\epsilon$).

A simplifying assumption is that the energy-momentum dependence of the phase-shift is approximately Lorentz invariant.
Introducing the Mandelstam variable $s=\omega^2-p^2$ one approximates
\be \delta_M (\omega,{\bf p}) \simeq \delta_M (\sqrt{\omega^2-{\bf p}^2},0) = \delta_M(\sqrt{s}) \ . \ee
We have numerically checked the validity of this approximation, except for small deviations due to the Landau cuts
at finite momentum. The behavior of $\delta_M(\sqrt{s})$ is directly linked to the analytical structure of the $q-\bar{q}$ polarization function along the real energy axis.
In particular it is sensitive to the presence of mesonic bound states (where the phase-shift jumps from 0 to $\pi$),
and to the appearance of the unitary and Landau cuts (the latter at nonzero temperatures for unequal quark masses~\cite{Kunihiro:1991hp,Yamazaki:2013yua}). 
Detailed discussions on the analytical structure of the polarization function can be found in Refs.~\cite{Hufner:1994ma,Zhuang:1994dw,Yamazaki:2012ux, Wergieluk:2012gd, Blaschke:2013zaa}.

Integrating Eq.~(\ref{eq:omega0}) by parts, and neglecting the
vacuum contribution (``no-sea approximation'')~\cite{Blaschke:2013zaa} one finally obtains the final formula~\cite{Zhuang:1994dw,Hufner:1994ma,Blaschke:2013zaa}
\begin{widetext}
 \be \Omega^{(0)}_{M} (T,\mu_M)  = -\frac{g_M}{8\pi^3} \int dp p^2 \int ds \frac{1}{\sqrt{s+p^2}} \ \left[  \frac{1}{e^{(\sqrt{s+p^2}-\mu_M)/T}-1}
+ \frac{1}{e^{(\sqrt{s+p^2}+\mu_M)/T}-1}  \right]  \ \delta_M (\sqrt{s};T,\mu_M)    \ . \ee
\end{widetext}

\section{Pressure of a quark---Polyakov loop---meson mixture}

  We present our results for the total pressure of the system at zero chemical potential,
\be P(T)= - [ \Omega_{\rm{PNJL}} (T,\mu_i=0) - \Omega_{\rm{PNJL}} (T=0,\mu_i=0)] \ee
with the grand-canonical potential at ${\cal O}(1/N_c)$
\be \Omega_{\rm{PNJL}} (T) = \Omega_q^{(1)} (T) + \sum_M \Omega^{(0)}_M (T) + {\cal U}_{\rm{glue}} (T;\Phi,\bar{\Phi}) \ . \ee

\begin{figure}[htp]
\begin{center}
\includegraphics[width=0.45\textwidth]{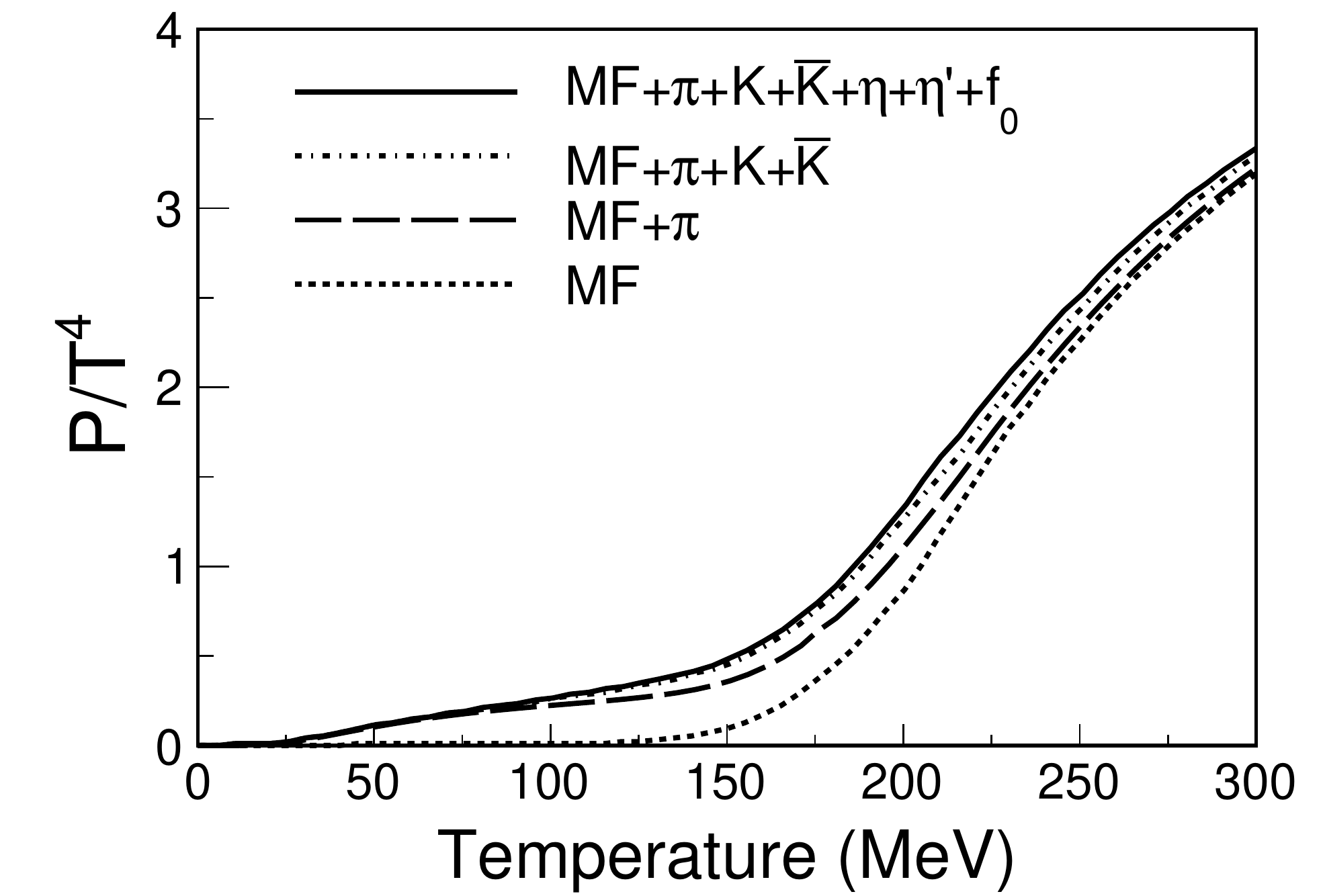}
\includegraphics[width=0.45\textwidth]{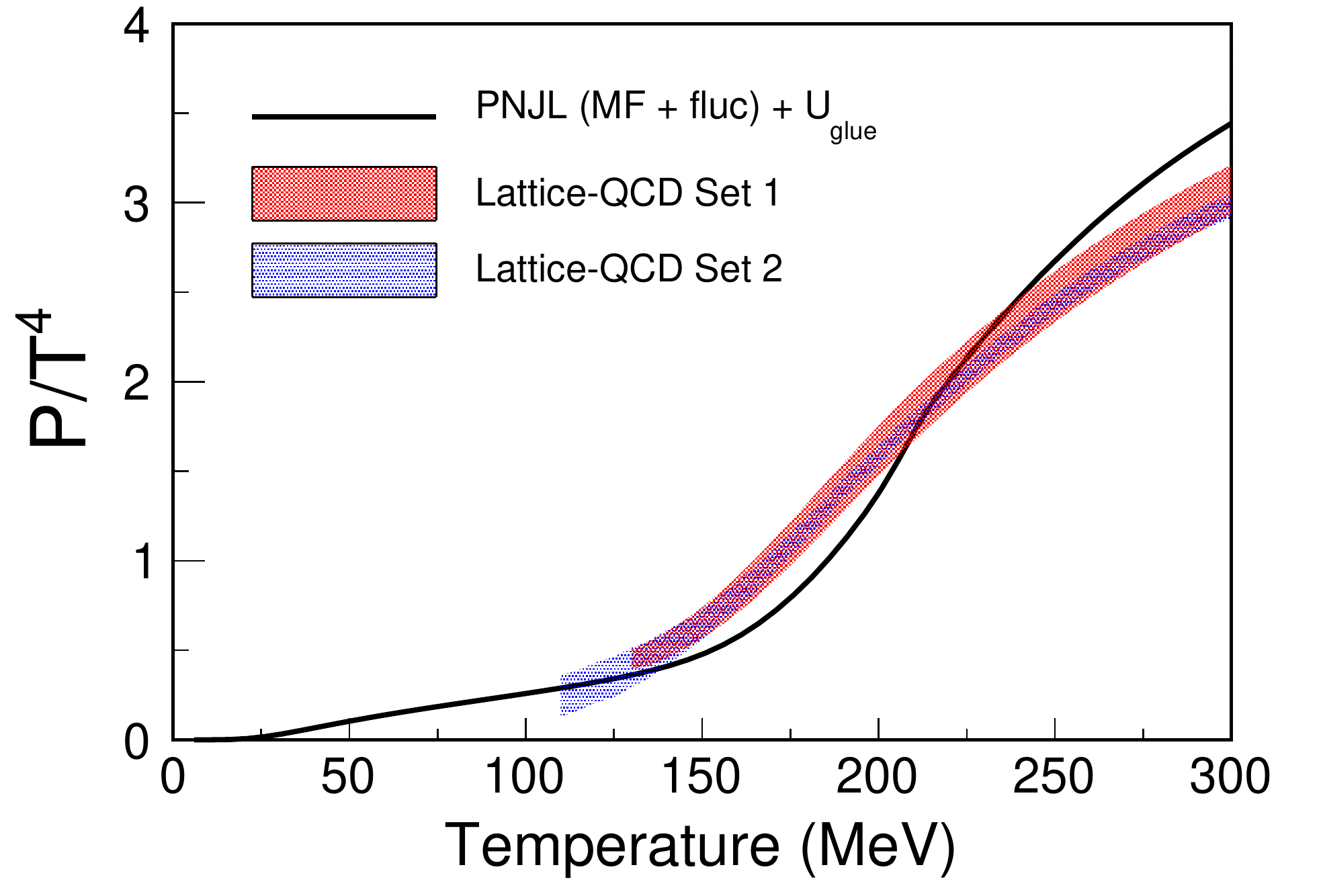}
\caption{\label{fig:Pcomparison} Pressure of the PNJL model at vanishing chemical potential with the effect
of mesonic states: $\pi,K,{\bar K},\eta,\eta',f_0$. Upper panel: Progressive contributions. Lower panel: Comparison with lattice-QCD calculations of Refs.~\cite{Borsanyi:2013bia,Bazavov:2014pvz}.}
\end{center} 
\end{figure}

   In the upper panel of Fig.~\ref{fig:Pcomparison} we show the progressive addition of different mesonic
states: 1) mean-field result (gas of quark-antiquarks and static gluons) 2) effect of pionic correlations
 3) kaon-antikaon contribution 4) all scalar and pseudoscalar mesonic states with a vacuum mass $< 1$ GeV ($\pi,K,\bar{K},\eta,\eta',f_0$). 
Notice that the more massive states still have an appreciable contribution at high temperatures. This is related
to the fact that in the PNJL model the melting (or Mott) temperatures of mesons are large in comparison with the
transition temperature~\cite{Torres-Rincon:2015rma}. Although they eventually melt down
above $T_c$ and only appear as scattering states, they still have tiny contribution of a few \% of the total pressure
(which becomes sizable when it is multiplied by the total number of states e.g. $\sum_M g_M=18$ for all $J^\pi=0^\pm$ states).
This effect needs to be readdressed in the future. In the lower panel we compare our outcome with two recent data sets from lattice-QCD results at zero chemical potential, Set 1 is taken from Ref.~\cite{Borsanyi:2013bia} and Set 2 from Ref.~\cite{Bazavov:2014pvz}.

  The results are quite satisfactory given the fact that for the application to the PNJL model at mean field, we
have only modified a single parameter $T^{cr}$ from the parametrization given in Ref.~\cite{Haas:2013qwp}. 
In addition, the contribution from the mesonic fluctuations carries no new parameters. Around $T_c$ the
underestimation of the total pressure can be systematically cured by adding more hadronic states (other mesons and even baryons).  

  Given the reasonable good comparison with lattice-QCD data we plan to explore the QCD phase diagram at finite
chemical potential. As opposed to current lattice-QCD calculations, the PNJL model allows for a straightforward
extension to this case. 

\acknowledgments

 We thank D. Blaschke for useful discussions and K. Yamazaki for email exchange. This work has been funded by the 
programme TOGETHER from R\'egion Pays de la Loire and EU Integrated Infrastructure Initiative
HadronPhysics3 Project under Grant Agreement n. 283286. JMTR also thank funded from the Spanish 
Ministerio de Ciencia e Innovaci\'on under contract FPA2013-43425-P.

\end{document}